# Separate-scan atomic force microscope for fast infrared scattering-type scanning near-field optical microscope

Yusuke Sakiyama[1, a)*], Emanuel Pfitzner[1, b)], Santiago H. Andany[2, c)], Georg E. Fantner[2], Joachim Heberle[1]

[1] Department of Physics, Exp. Molecular Biophysics, Freie Universität Berlin, 14195 Berlin, Germany

[2] Laboratory for Bio- and Nano-Instrumentation, Swiss Federal Institute of Technology Lausanne (EPFL), Lausanne 1015, Switzerland

[a)] current address: Nearfield Instruments B.V. Vareseweg 5, 3047 AT, Rotterdam, The Netherlands

[b)] current address: Attocube system AG, Egelfinger Weg 2, 85540 Munich-Haar, Germany

[c)] current address: Nanosurf AG, 4410, Liestal, Switzerland

* corresponding author: yusuke.sakiyama@gmail.com

## Abstract

Pseudo-heterodyne scattering-type scanning near-field optical microscopy (sSNOM) is applied in the mid-infrared region to detect the chemical composition of biomolecules on the nanoscale. However, the application of sSNOM in molecular biology has been limited to static images in air. Recently, bottom illumination sSNOM (BI-sSNOM) was developed for operation in water. Yet, the scan rate of sSNOM remains a bottleneck to record protein structural changes in aqueous solution on the seconds time scale. We designed an optical and mechanical system consisting of a separate scan high-speed atomic force microscope (HS-AFM) coupled to the BI-sSNOM optics. The designed AFM scanner has a mechanical bandwidth of ca 70 kHz along the Z-axis, and ca 6 kHz along the XY-axis, equivalent to the sample scanning HS-AFM. The AFM performance is demonstrated by imaging actin filaments. The optical design is validated by sSNOM experiments on purple membranes and microtubules.

## I. Introduction

Optical microscopes are the standard technique to visualize cells and organelles. Super-resolution optical microscopy, such as stimulated emission depletion (STED) or scanning near-field optical microscopy (SNOM), offer sub-diffraction spatial resolution, extending their applications to smaller biomolecules (e.g. the eightfold symmetry of the Nuclear Pore Complex has been resolved by STED and stochastic super-resolution microscopy) [1–4]. However, most of these techniques require fluorophores or antibodies which could induce undesirable effects on proteins. In contrast, infrared radiation is absorbed by chemical bonds. Despite its high specificity, the wavelength of IR radiation limits the spatial resolution of conventional far-field IR microscopy techniques due to diffraction [5]. To overcome the diffraction limit, a probe of atomic force microscope (AFM) is used as a local scattering probe which utilizes the near-field effect [6,7]. This infrared scattering scanning near-field optical microscopy (sSNOM) is able to distinguish secondary structures of proteins, like α-helices from β-sheets in amyloid like fibrils [8], image purple membrane [9], lipid bilayers [10] and thin-sectioned cells at a spatial resolution of several tens of nanometers [11]. As another microscopic tool to resolve the topographic information of protein dynamics, high-speed atomic force microscopy (HS-AFM) has been developed [12]. Walking myosin V, structural rearrangement during catalytic cycle of Roterless F1-ATPase and the assembly of centriolar scaffold proteins have been observed by HS-AFM [13–15].

There are three technical difficulties to apply sSNOM to dynamics of biomolecules in water. First, mid-IR is strongly absorbed in water. Second, the tip impact perturbs conformations and dynamics of soft proteins. Third, the scan rate is too low to capture conformational changes of proteins. Bottom illumination sSNOM (BI-sSNOM) has recently been developed to enable pseudo-heterodyne sSNOM imaging in aqueous solution [16–18]. However, an sSNOM-compatible AFM has to be designed to drive the low spring constant and high frequency cantilever to reduce the impulse from the tip to proteins. To increase the scan rate, bandwidths of sSNOM optics and AFM mechanics need to be improved [19]. In this work, we redesigned an optomechanical system of HS-AFM and combined it with a pseudo-heterodyne BI-sSNOM (Fig. 1). While XYZ sample-scanning AFM is mostly employed for side illumination sSNOM, the Z-piezo blocks the beam path of the BI-optical system [20]. XYZ tip-scanning HS-AFM has previously been developed and coupled to fluorescent SNOM, however the pseudo-heterodyne detection was not used and laser beam tracking to the AFM probe, especially in the visible region, remained a complication [21,22]. A XY sample-scanning benefits from maintaining the AFM probe in the focal point. Z tip-scanning is advantageous in the bandwidth to drive only the cantilever instead of the Si prism. Hence, we developed an optomechanical design of a separate scan HS-AFM, with XY sample-scanning and Z tip-scanning.

## II. Overall configuration and the optical design

1. AFM and the bottom illumination optical design

The AFM is designed to be coupled to a bottom illumination microscope. The AFM is placed on the optical breadboard, together with the pseudoheterodyne sSNOM detection system described in previous studies [18,23]. A quantum cascade laser (QCL) (model number 21060, Daylight solutions, USA) is used as a mid-infrared light (mid-IR) source with an emission tunable in the range of 5.78 - 6.39 μm. The configuration to focus mid-IR on the tip of the cantilever is as follows, mid-IR guided into a Schwarzschild objective lens (SOL) (LMM-40X-P01, Thorlabs, USA) is focused on the surface of a hemispherical solid immersion lens (SIL), the tip of sSNOM probe is placed on this focal point (Fig. 1). The near-field interaction between the tip and the sample enhances sidebands of higher harmonics n of the AFM oscillation frequency Ω. On the other side of beam splitter (BS), the Michelson interferometer arm oscillates at the frequency M (Fig. 1). Interference of the tip scattered and reference light introduces side-bands around the harmonics nΩ at nΩ + mM. Sidebands were extracted using a lock-in amplifier (LIA, HF2LI, Zurich Instruments, Switzerland). The lock-in-amplifier was controlled by a Labview application to map near-field signals as previously reported[18].

2. The validation of the optical system by simulations

The amide-I band or ca 6 μm wavelength are typically used for our sSNOM experiments. The angle of incidence from the SOL is between 14° and 30° (Fig. 2a). Since the critical angle between Si and air is 17°, we estimate that the incident light is divided into approximately 25% light refraction and 75% evanescent field (21.6° at Si-water).

Positioning the sSNOM tip in the focal point is a major technical challenge of our design. To enable efficient alignment, the IR intensity distribution at the silicon interface has to be intuitively understood. We simulated the electric field (E-field) propagating in the Si prism using the finite-difference time-domain (FDTD) method (Lumerical, Ansys, Inc. USA) and compared it to experimental data (Fig. 2b, c). Gaussian sources at 6 μm wavelength were placed on the bottom of the simulation volume and a Pt tip is placed on the Si surface in this study (supp. Fig.1, a1, b1). We investigated the Z-component of the E-field at the Si/air interface to find the optimal conditions for the near-field since the Z-component induces stronger near-field around the tip. The sSNOM amplitude image from the experiment shows two lobes similar to the Z component of the E-field ($E_z$) from FDTD (Fig. 2 b,c). The phase shift is seen at the separation of two lobes of amplitude in FDTD (supp. Fig. 1, a3, b3). While the Si prism is stationary in the FDTD calculation, it moves in the experiment, which leads to discrepancies in the phase maps. Due to the relatively high refractive index of silicon (3.47 for 6 μm wavelength), lobes are smaller than the waist size of 6 μm wavelength light in air (Fig. 2b, c). The motion of the prism leads to the focal point moving while XY scanning, this limits the maximum scan range (ca 1 μm x 1 μm) within the lobe (Fig. 2b).

The amplitude of $E_z$ is maximum at around 20 - 25° in air and 25 – 30° in water. It declines with a larger angle of incidence as calculated by FDTD (Supp. Fig. 2, 3). In previous studies using a ZnSe prism, the critical angle was used for the incident light [26,27]. The near-field associated scattering is increased when the p-polarization is aligned to the tip-sample dipole [28]. Considering the p-polarization and the E-field intensity, the optimal angle of incident light for the silicon/water interface is estimated to be 20 - 35° based on these calculations.

## III. Scanner

1. Z tip scanner

The HS-AFM optical head developed in the Fantner group [29–31] was further modified as a Z tip-scanner in this work. A piezo actuator (PA3CE, Thorlabs, USA) with a maximum travel range of 2 μm is fixed on the AFM head (fig. 3b). The tilt angle of an aluminum cantilever stage (tilt-stage) is 11° (fig. 3b). This ensures a perpendicular incidence of a readout laser to the cantilever. To seal the piezo from liquid, a silicone elastomer (SF00-2k silikon, silikon fabrik, Germany) fills the gap around the piezo. A photothermal laser drive and a conventional shaker piezo drive are equipped to drive the cantilever [15].

The cantilever is fastened on the tilt-stage using adhesive such as paraffin wax or nail polish. Although these adhesives ensure a tight fixation, there are two drawbacks. Firstly, the potential absorption of mid-IR radiation by adhesive when adhesive is immersed in aqueous solution. Secondly, the mechanical durability of the tilt-stage might be reduced, as undesirable force might be applied to the Z-piezo and the tilt-stage during the removal of adhesive (e.g. it takes more than 5 minutes to dissolve cured adhesive in acetone). Therefore, a mechanical clamp to hold the cantilever is preferred. The advantage of a horizontal beam clamp over a vertical clamp has been reported by Fukuda [32]. In accordance, the horizontal beam clamp is also used for our system. Screws and the beam are tilted by 11°, so that the cantilever chip body is pressed perpendicularly. The beam clamp has a convex to hold the chip body and slits move as flexures to ensure a high frequency response (Fig. 3c). Finite element analysis (FEA) was performed using SolidWorks software (SolidWorks Corp., USA) to understand the vibration modes of the clamping mechanism. Stainless steel (X8CrNIS18), with elastic modulus of 200 GPa, a Poisson's ratio of 0.28, and a density of 7900 kg/m$^3$, were used for simulations. The first mode frequency of the beam clamp estimated by FEA is 180 kHz (Fig. 3d). The frequency response was experimentally determined using a vibrometer (Picoscale, SmarAct GmbH, Germany) and a piezo driver (PD200, PiezoDrive, Australia). With wax, peak 2 has a resonance frequency of ca 102 kHz and a displacement amplitude of 36.6 nm (Fig. 4). With the beam clamp, both peaks are considerably shifted to lower frequencies with a frequency of ca 65 kHz of peak 2 and an amplitude of 106.1 nm. The lower resonance frequency and the higher peak amplitude are caused by preload of the beam clamp. The phase delay starts at about 70 kHz, which is close to the frequency of peak 1 (Supp. Fig. 4b). The phase delay was measured using a custom-built interferometer because the SmarAct vibrometer used doesn't record the phase delay.

The reduction of an undesirable vibration (peak 1) by a counter balance Z-piezo is a commonly used technique in HS-AFM (Fig. 4) [12,22,30,32,33,34]. Due to the working distance of the focusing lens and the limited space of the housing, it is not possible to mount the counter Z-piezo in our AFM head. An alternative way to overcome such coupling mechanisms is the use of model based or data driven controls [35].

For our application, the main resonance frequency of ca 65 kHz with a beam clamp, is sufficient for the fast sSNOM measurement. The state-of-the-art sSNOM pixel rate is a few hundred hertz, and the pixel rate required for fast sSNOM is up to ten kilohertz. When the maximum speed performance of AFM is required, the cantilever is fastened with adhesive.

2. XY Sample scanner

The challenge in designing an XY sample scanner loaded with an optical prism is to achieve a bandwidth beyond 2 kHz. We used a symmetric design with multiple flexures to hold a sample and induce movement for the scanning direction, which is based on the configuration by Fantner [36]. Although the depth of the flexures or the thickness of the XY-scanner plate is a crucial factor for the mechanical stability, it is limited by the opening angle of the bottom optical window. The optimal thickness of the XY-scanner or the flexure depth is 4 mm, resulting in the bottom opening angle of 36.9°, which covers the maximum incident light angle from SOL of 30° (Fig. 2a).

The flexure must be flexible enough for sufficient displacement while maintaining high rigidity to achieve a high resonance frequency. FEA simulations were performed to optimize the flexure length. A titanium alloy (Ti-6Al-4V) with an elastic modulus of 104.8 GPa, a Poisson's ratio of 0.31 and a mass density of 4429 kg/m$^3$ were used for simulations. The maximum flexure length is 5 mm due to the XY scanner housing. We tested FEA simulations on the scanner with different lengths of the flexure (2-5 mm). A scanner with flexure length of 2mm, 3 mm, 4 mm and 5 mm are called flexure-2, flexure-3, flexure-4 and flexure-5 respectively in this paragraph. FEA calculations were performed for the scanner loaded with a silicon prism, a spring on both sides instead of a piezo due to the limited functionality of the SolidWorks software. The static displacement with 100 N of applied force and the vibration mode were simulated by FEA (fig. 5b,c). For instance, the frequency of mode 1 at 3.06 kHz with flexure-5 and 4.8 µm displacement with flexure-2 are not sufficiently high or large (table 1). We decided flexure-3 and flexure-4 to be manufactured by electric discharge machining and tested these experimentally. Grade 5 titanium was used for its high fatigue limit property [30].

Stack piezos (PC4WL, Thorlabs, USA) with 235 kHz free resonance frequency are mounted on the X- and Y-scanner. Stainless steel springs (spring constant of 31.8 N/mm, D117-E, Gutekunst Federn, Germany) are inserted for preloading on the counter side of the piezos (fig. 5a). Gap spaces are filled with the elastomer silicone (SF00-2k silikon, silikon fabrik, Germany) to passively damp undesirable vibrations. The vibration mode 2 is a twisting motion occurred over 10 kHz, which is well above the X-scan rate and is suppressed by the elastomer (Fig. 5 c, table1). The frequency responses were measured using the smarAct vibrometer. Flexure-3 has resonance frequencies at 10.1 kHz and 13.1 kHz. Flexure-4 has resonance frequencies at 9.3 kHz and 10.5 kHz (Fig. 6a). The phase delay of the frequency response starts at ca 8 kHz for flexure-3 and at ca 6 kHz for flexure-4 (Fig. 6b). The phase delay was measured using a custom-built interferometer separately from the amplitude measurements by the smarAct vibrometer. The resonance frequency of flexure-4 is high enough for the highest line rate required for HS-AFM. Flexure-3 has undesirable small spikes and a lower peak than flexure-4, caused by the excess stiffness of flexure-3 (Fig. 6a). Flexure-4 is used for further AFM and sSNOM experiments as it has better stability and higher displacement efficiency.

### IV. AFM and sSNOM image

1. AFM images

The usage of short, high frequency cantilevers enables a high mechanical bandwidth. Typically used large tapping mode cantilevers such as Arrow-NC (Nanotools GmbH, Germany) or RTESP-300 (Bruker AFM Probes, USA) limit the mechanical bandwidth to less than 2 kHz in air [29]. To demonstrate the separate-scan AFM generate images at an adequate scan rate for fast sSNOM, actin filaments were imaged in air using an ultra-short lever with a carbon tip (USC-F2-k3, Nanotools, Germany)(Fig. 8 a, b, c). Globular actins (Hypermol e.K. Germany) were polymerized prior to AFM experiments. Silane functionalized mica is prepared, such that actin filaments are immobilized on the surface [37]. The mechanical crosstalk between the z-scanner and the shaker piezo cause noise in the image captured with 60 lines/s (Fig. 7c). The photothermal drive provides the solution to this crosstalk [15,31,38]. An image acquired with photothermal tapping using a Fast-scan A lever (Bruker AFM Probes, USA) (Supp. Fig. 6c) shows less noise than other images acquired with shaker tapping at the same line rate of 118 lines/s (Supp. Fig. 6b, c, d).

The bandwidth of the system is limited by the Nanoscope IIIa controller (Bruker Corp., USA) (supp.fig. 5), however the separate scan AFM recorded images at a sufficient scan rate to increase the scan rate of the state-of-the-art sSNOM (Fig. 7, supp.fig. 6a).

2. sSNOM experiments in air

We performed sSNOM experiments of PM on a Si prism to verify the optical design. Figure 8 a-c shows the optical image (n = 3) at 1.0 lines/s, with the QCL tuned to 1659 cm$^{-1}$. Same PMs are found in an AFM topography and optical images (panels b and c for amplitude and phase imaging, respectively). A line-by-line linear fit is subtracted from the optical phase map while a median filter further suppresses the horizontal noise. The optical maps and data corrections were processed using Python code. The optical power of the collimated light used for the bottom illumination is approximately 2-3 mW. The optical image proves that the sSNOM optical design fits well into the mechanical design of separate-scan type HS-AFM configuration. For faster optical imaging, microtubules were used as a test sample because of the higher signal from larger protofilaments. The microtubule polymerization assay was performed using porcine brain tubulins (Cytoskelton, Inc., USA) following previous studies[39]. Microtubules were cross-linked by glutaraldehyde after polymerization. Microtubules were deposited on the silanized silicon surface. The sSNOM experiment was performed using an Arrow-NCPt lever. Optical amplitude and phase maps show a microtubule at ca 13 seconds per frame, 9.86 lines/s or ca 770 μs per pixel (Fig. 8, d-f). The effective size of the sSNOM probe estimated from the optical phase map was 42.5 nm (supp. Fig. 9).

3. sSNOM experiments in water

Arrow-NCHPt or 240AC-GG (OPUS, Mikro Masch, Bulgaria) lever used in previous studies for liquid sSNOM operation are not suitable for fast scanning in water as described above [16,18]. Thus, we selected the fast-scan B lever (FS-B, Bruker, USA) for platinum coating for sSNOM application in water. The mechanical bandwidth of a platinum coated FS-B (FSB-Pt) and a standard sSNOM lever (arrow-NCHPt) were ca 8 kHz and ca 250 Hz, respectively (Supp. Fig. 10).

sSNOM experiments of microtubules in tubulin buffer (80 mM PIPES pH 6.9, 2 mM $MgCl_2$, 0.5 mM EGTA) were performed using the designed separate scan AFM, the BI-sSNOM optics, an FSB-Pt lever and the reference arm (M = 1100 Hz). Due to the depolymerization of microtubules and ca 30 minutes of the duration of the optical alignment, cross-linked microtubules used for sSNOM experiments in air were rehydrated in tubulin buffer for these validation experiments. Optical images were recorded at ca 770 μs per pixel, 9.86 lines/s or ca 13 seconds per frame (Supp. Fig. 11). Optical images of scattered light from the sSNOM tip show filament shape of microtubule in aqueous solution.

## V. Conclusion and outlook

An optomechanical system of separate scanning HS-AFM coupled with bottom illumination pseudoheterodyne sSNOM was designed and experimentally tested. Without tracking of the cantilever by the AFM readout laser at high frequency, AFM imaging is enabled by this separate scanning system. The mechanical bandwidth of the Z-scanner is ca 70 kHz. The XY-sample scanner with the Si prism has a resonance frequency of over 6 kHz. AFM images were examined with a pixel rate of up to ca 30 kHz and with the spatial resolution to visualize separations of actin filaments. The advantage of the photothermal drive for tip-scan AFM over the shaker drive is demonstrated by the lower noise level at the pixel rate above 10 kHz.

Fast sSNOM experiments were performed in aqueous solution using the designed AFM, the BI-sSNOM optics, the shorter sSNOM lever and the reference arm. The optical images of the light scattered from the sSNOM tip demonstrate the filament shape of microtubules with a frame rate of 13 seconds. Increasing the signal-to-noise ratio will allow a further increase of the scan rate [40]. By synchronizing the motion of the focal spot and the AFM XY-scanning trace by the galvo scanning mirror, a larger sSNOM scan area will be covered, as previously reported [18,24]. Optimizing the incident angle of the bottom illumination, and a plasmon nanofocusing by a slotted sSNOM probe are considerable options [41]. The broadband infrared spectrum using fast sSNOM in an aqueous solution is the forthcoming advancement. The effectiveness of the peak-force tapping for near-field signals by sSNOM and protein observations by HS-AFM has been demonstrated independently and might be an alternative strategy for near-field modulation instead of amplitude modulation AFM [15,42]. Fast sSNOM with peak-force tapping has the potential to probe proteins in aqueous solution mechanically gently with the optimized near-field effect [43].

As a comparable technique to sSNOM, the thermal expansion of biomolecules after infrared irradiation is measured by AFM cantilever deflection (AFM-IR). The applicability of AFM-IR for cells, bacteria [44–46], organic materials and fibrillar amyloids has been demonstrated [47]. By synchronizing laser pulses with a duration of tens of nanoseconds and the peak force between the tip and the sample, peak-force AFM-IR offers several advantages, including a high-precision control, a lower infrared irradiation power, lower tapping force, and Z-drift compensation [48]. However, peak-force AFM-IR has a slower detection rate than amplitude modulation tapping AFM-IR [49]. The decay time of thermal expansion [50] and the resonance frequency of the cantilever limit the temporal resolution of AFM-IR. Another drawback of AFM-IR is the inability to extract the optical absorption independent of the mechanical amplitude. Hence, we consider that a light scattering technique like sSNOM is more suitable to couple with HS-AFM.

We envision the fast sSNOM to be applied to a single large protein or a giant enzyme to understand orientation changes of domains by probing the Amide-I, II band [51], in addition to conformational changes imaged by HS-AFM [13,52,53].

SUPPLEMENTARY MATERIAL

Details of simulation methods, the frequency response of the Z-scanner, the schematic of the feedback loop, imaging data for supporting the main text, a cross-section view of the optical phase map.

ACKNOWLEDGMENTS

We thank Ramona Schlesinger and Kirsten Hoffmann (FU Berlin, Genetic Biophysics) for providing purple membranes. We thank Mathias von Borcke (FU Berlin, electronic workshop) for helping with electronics. We thank Gabriela Luna Amador and Stephanie Reich (FU Berlin, experimental solid-state physics) for using the Ansys Lumerical software. We thank Gada Hweidi and Marc Benjamin Hahn (Federal Institute for Material Research and Testing, BAM, Berlin) for using their vibrometer. This work was funded by the Alexander von Humboldt Foundation (1214529 - JPN - HFST-P to Y.S.) and the Investionsbank Berlin (Provalid, VAL20/2023 to Y.S.).

AUTHOR DECLARATIONS

Conflict of Interest

Intellectual property

{Y.S., and J.H.} has Patent {DE102024103264 {pending}}.

Autor Contributions

**Yusuke Sakiyama**: Conceptualization (equal); Funding acquisition (lead); Investigation; Methodology (lead); Visualization; Writing – original draft; Writing - review & editing (equal); Project administration (equal). **Emanuel Pfitzner**: Conceptualization (equal); Methodology (supporting); Writing – review & editing (equal). **Santiago H. Andany**: Conceptualization (equal); Methodology (supporting); Writing – review & editing (equal). **Georg E. Fantner**: Conceptualization (equal); Methodology (supporting); Writing – review & editing (equal). **Joachim Heberle**: Funding acquisition (supporting); Writing – review & editing (equal); Project administration (equal).

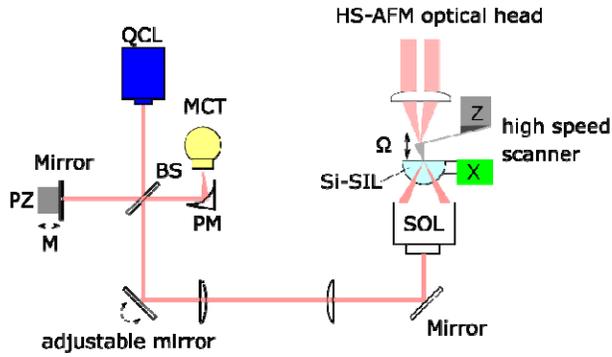

Fig. 1 Schematic of the pseudoheterodyne sSNOM coupled to the mechanics of HS-AFM. IR-laser illuminates Si prism (Si-SIL) from bottom through Schwarzschild objective lens (SOL). Light scattered from sSNOM tip is collected by SOL and returned to the beam splitter (BS), focused on to the marcury-cadmium-telluride detector (MCT) by the parabolic mirror (PM). The optical phase in the reference arm is modulated by piezo (PZ) driven mirror.

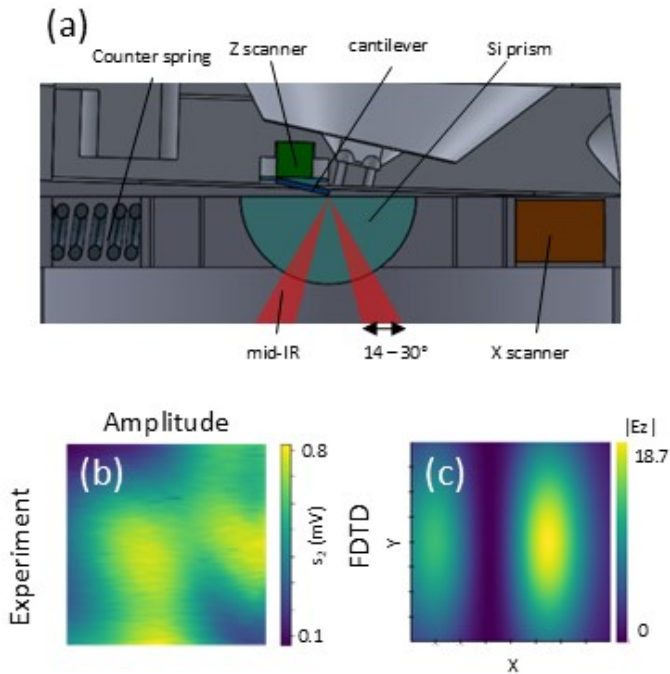

FIG. 2. Comparison of experimental near-field signal and calculated Z-vector of E-field. (a) The section view of the CAD design of the bottom illumination and AFM scanner. (b) sSNOM amplitude image (n=2). 5 lines/s, 4 x 4 µm² (maximum scan range of XY scanner). (c) amplitude of Z-vector of E-field (Ez). Wide field of view is presented in supp. Fig. 1a-2.

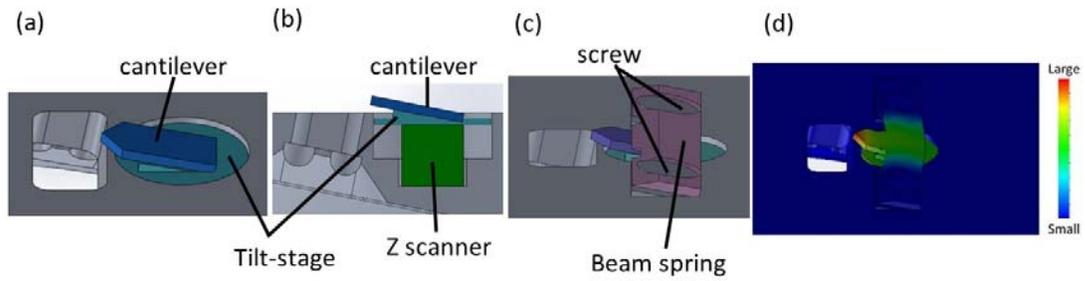

FIG.3. Schematic diagram of Z tip scanner. (a) isomeric view (b) side view (c) isomeric view with mechanical clamp holder (d) 1$^{st}$ vibration mode of beam spring calculated by FEA. (F = 180kHz)

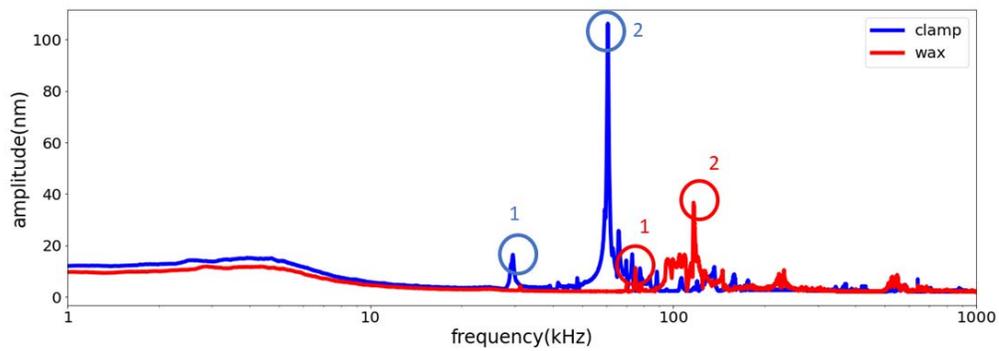

FIG. 4. Frequency response of Z tip scanner. Amplitude sweep measured using a vibrometer (SmarAct). 5 V is applied to Z piezo. The cantilever is fixed with a mechanical clamp or wax.

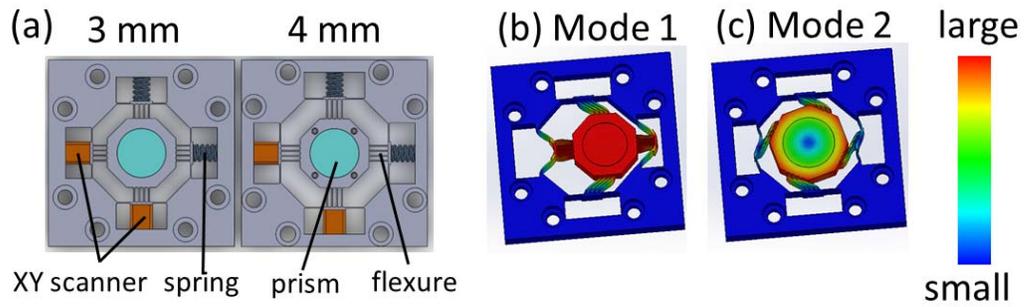

FIG. 5. (a) XY sample scanner. Flexure length is 3 mm and 4 mm, respectively. (b), (c) Vibration mode of the XY scanner is simulated by FEA.

| FEA \ flexure length | 2 mm | 3 mm | 4 mm | 5 mm |
|---|---|---|---|---|
| Frequency Mode 1 | 14 kHz | 5.39 kHz | 5.04 kHz | 3.06 kHz |
| Frequency Mode 2 | 26.4 kHz | 14.17 kHz | 10.85 kHz | 10.2 kHz |
| Static displacement by 100 N | 4.8 μm | 34 μm | 36 μm | 98 μm |

Table 1. FEA results of frequency mode and static displacement for each flexure length.

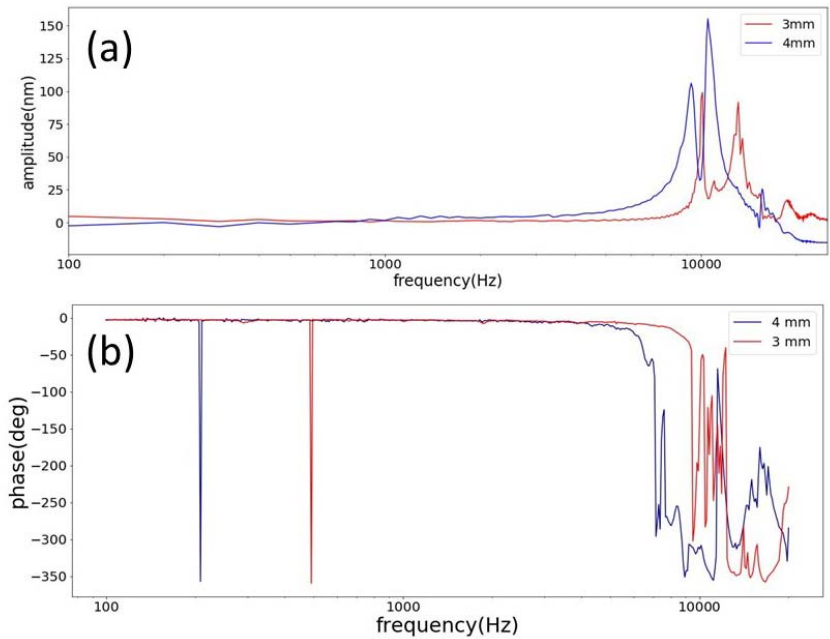

FIG.6. Frequency response of the X-scanner. (a) Amplitude of frequency response measured using vibrometer (SmarAct). Comparison of scanner with 3 mm and 4 mm flexure. 1 V applied to X-piezo. (b) Phase delay of frequency response measured using a custom-built interferometer.

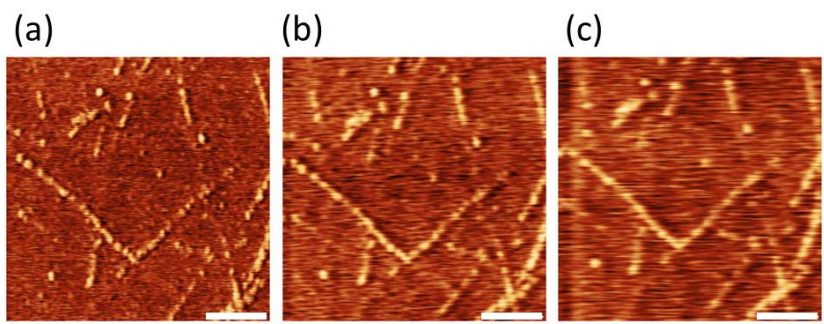

FIG.7. AFM images of actin filaments in air. Scale bar 200 nm. (a) 20 lines/s (b) 40 lines/s (c)60 lines/s.

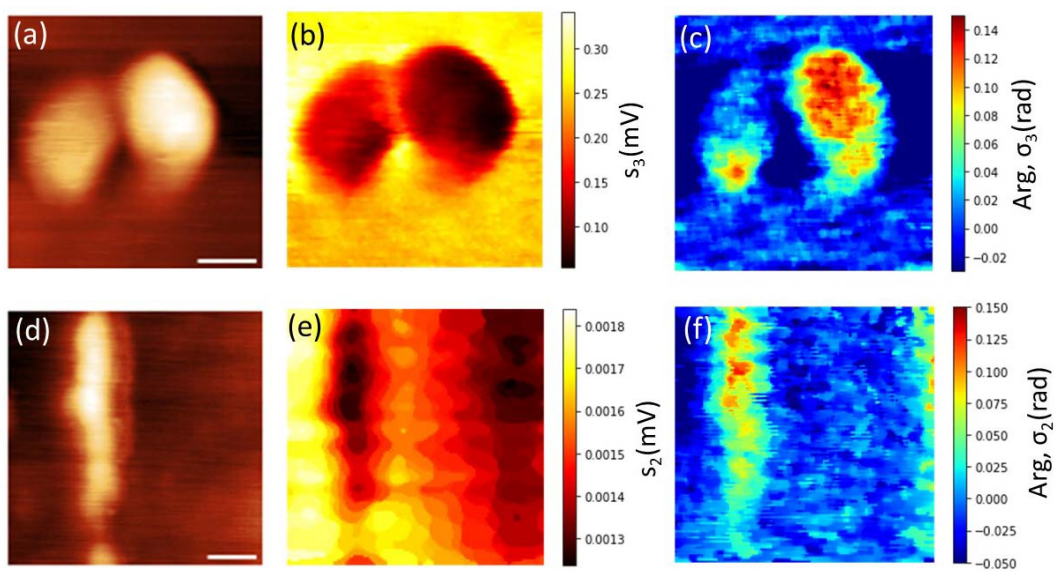

FIG.8. sSNOM experiments in air. (a-c) Purple membrane. Wavenumber 1660 cm$^{-1}$. 1 lines/s (a) AFM topography, scale bar 200 nm (b) Optical amplitude image. (c) Optical phase image. (d-f) microtubule. Wavenumber 1660 cm$^{-1}$. 9.86 lines/s (d) AFM topography, scale bar 200 nm. (e) Optical amplitude image. (f) Optical phase image.